\documentclass[prl,amsmath,amssymb, superscriptaddress,twocolumn]{revtex4-1}

\usepackage{graphicx}
\usepackage{dcolumn}
\usepackage{bm}
\usepackage{color}
\usepackage{xcolor}
\usepackage{hyperref}
\usepackage[normalem]{ulem}

\usepackage[mathlines]{lineno}

\begin{document}


\title{Friction-controlled reentrant aging and fluidization in granular materials}

\author{Ye Yuan}
\affiliation{Research Center for Advanced Science and Technology, The University of Tokyo, 4-6-1 Komaba, Meguro-ku, Tokyo 153-8904, Japan}

\author{Walter Kob}
\email{walter.kob@umontpellier.fr}
\affiliation{Laboratoire Charles Coulomb, University of Montpellier and CNRS, 34095 Montpellier, France}

\author{Hajime Tanaka}
\email{tanaka@iis.u-tokyo.ac.jp}
\affiliation{Research Center for Advanced Science and Technology, The University of Tokyo, 4-6-1 Komaba, Meguro-ku, Tokyo 153-8904, Japan}
\affiliation{
Department of Fundamental Engineering, Institute of Industrial Science, The University of Tokyo,  4-6-1 Komaba, Meguro-ku, Tokyo 153-8505, Japan}

\date{\today}

\begin{abstract}
{\bf Granular materials densify under repeated mechanical perturbations, a nonequilibrium dynamics that underlies many natural and industrial processes. Because granular relaxation is governed by frictional contacts and energy dissipation, this aging behavior fundamentally differs from that of thermal glasses despite their apparent similarities.
Here, we uncover how friction controls the compaction dynamics of granular packings subjected to cyclic shear. Using discrete element simulations, we construct a dynamic state diagram as a function of strain amplitude and friction, revealing a rich interplay between jamming marginality, stabilization, and fluidization. We identify a friction-dependent crossover strain that separates aging and fluidized regimes, showing reentrant, non-monotonic behavior: Increasing friction first suppresses fluidization, then promotes it through smooth, creep-like rearrangements. This transition is marked by a shift from intermittent, avalanche-like rearrangements to continuous, diffusive motion. Our findings demonstrate that friction exerts a dual role in granular aging --- both stabilizing and fluidizing --- thereby uncovering the fundamental nonequilibrium mechanisms that govern compaction, rheology, and aging in athermal disordered systems. More broadly, our results reveal a general principle for how friction governs metastability and flow in athermal matter --- from granular and frictional colloids to soils and seismic faults --- linking microscopic contact mechanics to macroscopic dynamics.
}
\end{abstract}

\maketitle

Dense granular materials, such as soil, powders, and grains, are ubiquitous in natural and industrial processes. Though athermal and dissipative, they exhibit glass-like features such as constrained motion, slow relaxation, and dynamic heterogeneity~\cite{dauchot2005dynamical,keys2007measurement,watanabe2008direct,kou2018translational}. These behaviors, well known in thermal glasses~\cite{parisi2010mean,baule2018edwards,cubuk2017structure,nicolas2018deformation}, have motivated the application of glass physics concepts to granular dynamics. However, fundamental differences exist: Energy dissipation through particle deformation, interparticle friction, and surface roughness qualitatively alter relaxation pathways~\cite{kou2017granular,yuan2024creep,komatsu2015roles}. 
Unlike glasses, where thermal fluctuations drive relaxation toward equilibrium, granular dynamics is governed by external forcing and irreversible dissipation, giving rise to distinct out-of-equilibrium behaviors.

A key manifestation of nonequilibrium dynamics in both glasses and granular systems is \emph{aging}, i.e., the slow evolution of structural and dynamical properties following a perturbation. 
In glasses, aging corresponds to relaxation toward higher \emph{thermodynamic} stability~\cite{kob1997aging,tanaka2005kinetics,wang2024distinct,leishangthem2017yielding}, whereas in granular systems it often manifests as slow compaction under weak perturbations such as tapping~\cite{knight1995density,philippe2002compaction,richard2005slow,xia2015structural,gago2020universal}, cyclic shear~\cite{pouliquen2003fluctuating}, or compression--decompression cycles~\cite{kumar2016memory}. 
This compaction reflects a gradual evolution toward enhanced \emph{mechanical} stability. 
In both cases, the strength of external driving controls the rate and extent of relaxation. Yet in granular matter, dissipation makes aging fundamentally different: Unlike glasses, where thermal fluctuations enable reversible exploration of the energy landscape, it proceeds through irreversible rearrangements.

Although steady-state behaviors of driven granular systems are well studied --- both in terms of packing properties~\cite{schroter2005stationary,xia2015structural,yuan2021experimental} and relaxation dynamics~\cite{dauchot2005dynamical,keys2007measurement,reis2007caging,kou2017granular,zhao2022ultrastable,yuan2024creep} --- the nonequilibrium dynamics of mechanical aging under small, quasi-static perturbations remains poorly understood. 
In particular, the role of interparticle friction, essential for real grains, has not been clarified despite its fundamental impact on stability and rearrangements.

To advance on this question, we numerically investigate the compaction dynamics of disordered granular materials under quasi-static cyclic shear. 
By varying the shear amplitude $\Gamma$ and interparticle friction $\mu$, we construct a $\Gamma$--$\mu$ state diagram that reveals three regimes: (i) an \emph{elastic regime} at minimal $\Gamma$ and intermediate $\mu$; (ii) a \emph{mechanical aging regime}, with logarithmic-like compaction and heterogeneous dynamics --- \emph{intermittent} at low $\mu$, \emph{creep-like} at high $\mu$; and (iii) a \emph{fluidized regime}, where homogeneous steady-state dynamics emerge at large $\Gamma$, independent of $\mu$. 
Most strikingly, our results overturn the conventional view of friction as merely a stabilizing factor~\cite{silbert2010jamming}. 
Instead, we show that increasing friction first enhances stability, but then facilitates fluidization through creep-like micro-rearrangements enabled by abundant metastable states. 
This establishes a distinct physical principle: friction reshapes the nonequilibrium phase behavior of jammed assemblies in a manner fundamentally different from conventional glassy aging.

\section*{Results and discussion}

\subsection*{Simulation methods and protocol}

We employ the discrete element method (DEM) to simulate two-dimensional (2D) granular materials composed of frictional disks~\cite{silbert2001granular}. Particles interact via a linear spring-dashpot model with static friction, which accounts for both normal and tangential forces upon contact (see Supplemental Materials for further details). The system consists of 5,000 particles composed of a 50:50 bidisperse mixture with a size ratio of 1.4, which helps suppress crystallization and maintain a disordered structure. The radius of the smaller disks is set as the unit length.
To improve the accuracy of our results, we average most of them over 5--10 independent simulations.

Disordered jammed packings are generated using a quasi-static isotropic compression protocol, applied to systems with eight different friction coefficients, $\mu = 0$ (frictionless), 0.01, 0.03, 0.055, 0.1, 0.2, 0.3, and 1
, which allows us to systematically examine the impact of friction on the packing structure. We maintain a dimensionless pressure $P = 4 \times 10^{-3}$, close to the hard-particle jamming point, which yields packing fractions $\phi$ approximately 0.01 higher than the jamming onset defined by the limit $P \to 0$.

As shown in Fig.~\ref{Figure1}(a), our results reproduce the friction-dependent jamming transition observed in previous studies~\cite{song2008phase,silbert2010jamming}, and this behavior remains robust against small variations in $P$ (open and filled symbols in the graph).
Specifically, the packing fraction $\phi_0(\mu)$ and coordination number $Z(\mu)$ exhibit a marked decrease around $\mu \approx 0.2$, indicating that larger $\mu$ enable mechanical stabilization at lower $\phi$. This threshold separates the low-friction regime, where friction has little influence on mechanical stability, from the high-friction regime, where friction plays a dominant role in maintaining structural integrity. 
Interestingly, we find that with increasing friction the reduction in packing fraction from the frictionless jamming point, $\phi_{\rm J}(\mu=0) - \phi_0(\mu)$, is approximately proportional to the decrease in contact number, $Z_{\rm c}(\mu=0) - Z(\mu)$, see right ordinate of the graph. Here, $\phi_{\rm J}$ is the jamming packing fraction and $Z_{\rm c}$ is the isostatic contact number for frictionless particles. This proportionality indicates that the friction-induced lowering of jamming density is closely linked to a reduction in contact connectivity~\cite{song2008phase,silbert2010jamming}. 

Following the preparation of the jammed packings, we apply to the samples a quasi-static cyclic shear protocol, using a range of shear amplitudes $\Gamma$, as illustrated in Fig.~\ref{Figure1}(b). The fixed-pressure condition accommodates volume changes and enables systematic exploration of the system's response, avoiding complications associated with deep compression or excessively loose configurations that are present in constant volume simulations. This condition also closely reflects the experimental setup used to study the behavior of real dry granular materials. The cyclic shear setup thus provides a realistic framework to examine how driving amplitude $\Gamma$ and friction $\mu$ influence the aging dynamics of disordered granular systems. 
It is important to emphasize that our protocol fundamentally differs from continuous-energy-input perturbations, such as vibration or shear flow, in that it consists of discrete, quasi-static cycles that allow the system to relax between driving events.

\subsection*{Compaction process}

\begin{figure}
\centering
\includegraphics[width=8.5cm]{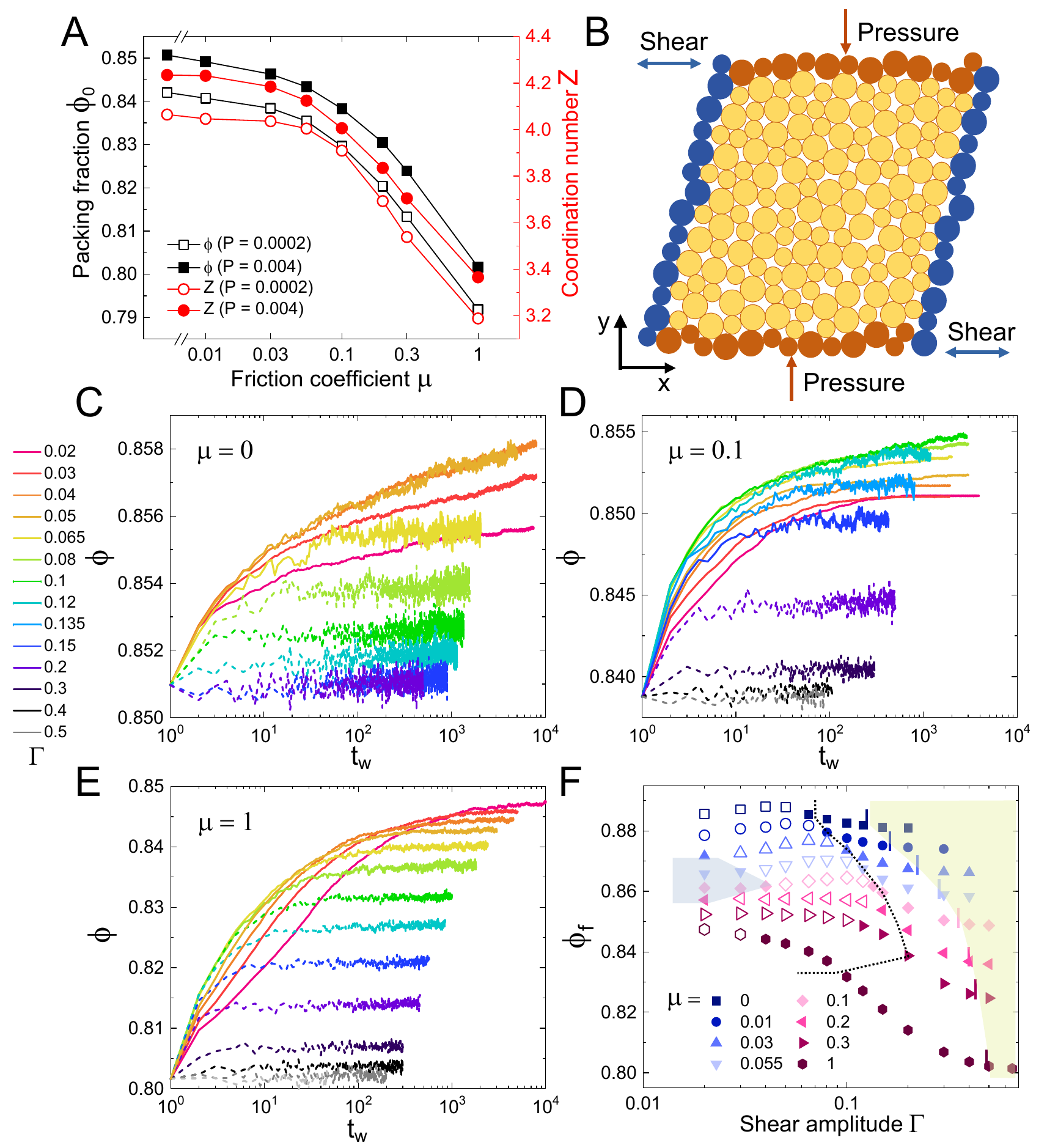}
\caption{
(a) Packing fraction ($\phi_0$) and coordination number ($Z$) as functions of the friction coefficient ($\mu$) obtained from a quasistatic compression protocol. The trends remains consistent under an imposed dimensionless pressure of $P = 0.004$ (solid symbols) compared to a much lower pressure of $P = 2 \times 10^{-4}$ (open symbols), approaching the hard-particle limit. 
(b) Simple shear setup: Side boundaries (blue) tilt to induce shear, while the top and bottom layers (red) shift (i) along the shear direction and (ii) adaptively perpendicular to maintain imposed pressure $P$.
(c-e) Packing fraction $\phi(t_w)$ during compaction for $\mu = 0$, $0.1$, and $1$, for various shear amplitudes $\Gamma$. 
Solid lines indicate a solid-like aging regime, while dashed lines denote a fluidized regime (see main text for definition). 
(f) Final packing fraction $\phi_{\rm f}$ vs. $\Gamma$ and $\mu$. Data for $\mu = 0 - 0.2$ are shifted upwards by 0.03, 0.025, 0.02, ..., 0.005 for clarity. Open symbols: Systems that are not yet in the steady state. Left shaded region: Elastic regime. Right shaded region: No compaction for $\Gamma > \Gamma_{\rm nc}(\mu)$; short lines mark $\Gamma_{\rm nc}(\mu)$. Dotted curves: Aging--fluidized boundary.
}
\label{Figure1}
\end{figure}

First, we describe the compaction process of the granular systems, prepared as described above, under cyclic shear as a function of the applied shear cycles $t_w$. Figures~\ref{Figure1}(c-e) show the evolution of the packing fraction $\phi$ with $t_w$ for three representative friction coefficients $\mu$; results for other $\mu$ values are provided in Fig.~S1. Independent of $\mu$, the systems undergo varying degrees of compaction, progressively evolving from their initial packing fractions $\phi_0(\mu)$ toward final values  that typically range from $\phi \approx 0.845$ for high friction ($\mu = 1$) to $\phi \approx 0.86$ for nearly frictionless systems ($\mu \approx 0$). Since higher friction leads to lower initial packing fractions $\phi_0(\mu)$, see Fig.~\ref{Figure1}(a), the extent of compaction naturally becomes more pronounced with increasing $\mu$. Whether the system continues to compact slowly or reaches a steady-state $\phi$ at the largest accessible $t_w$ depends sensitively on the interplay between $\Gamma$ and $\mu$. In the following we will study this dependence in more detail.

For small $\Gamma$, the packing fraction $\phi(t_w)$ exhibits a two-step evolution: An initial rapid increase during a transient regime, followed by a slower, logarithmic-like growth that can be identified as mechanical aging. A qualitatively similar behavior has been observed in compaction experiment under gentle tapping~\cite{knight1995density}. This two-step process mirrors aging in glasses, where local reorganization of the structure quickly enhances stability, and further densification proceeds slowly via intermittent or collective rearrangements~\cite{kob1997aging,tanaka2005kinetics,yunker2009irreversible,el2010subdiffusion}. The duration of the transient regime, denoted as $t^\ast$, generally increases with $\mu$, reflecting the growing strength of the mechanical contacts. In contrast, for larger $\Gamma$, the system rapidly attains a steady-state $\phi$ without undergoing slow aging. This indicates that at sufficiently large $\Gamma$, the system yields within just a few cycles and quickly enters a steady state.

As shown in Fig.~\ref{Figure1}\textit{D}, moderate $\mu$ and sufficiently small $\Gamma$ lead to a plateau in $\phi(t_w)$ after the initial transient, due to frictional locking~\cite{royer2015precisely,mao2024dynamic}. This plateau reflects the formation of a unique, friction-stabilized elastic configuration, i.e., an arrested configuration with minimal rearrangement during the cycling. We will discuss this absorbing state in more details in the following sections. 

Figure~\ref{Figure1}\textit{F} presents the final packing fraction $\phi_{\rm f}$, attained at our largest accessible $t_w$ (up to $10^4$) as a function of $\Gamma$.(Note that the curves for small $\mu$ have been shifted vertically). For $\mu \leq 0.1$, we observe a local maximum of $\phi_{\rm f}$ at an optimal strain amplitude $\Gamma_{\rm opt}$, at which the compaction process is most effective. The maximum packing is achieved at $\Gamma_{\rm opt} \approx 0.05$ for frictionless systems ($\mu = 0$) and at $\Gamma_{\rm opt} \approx 0.1$ for $\mu = 0.1$. In this range of $\mu$, moderate driving amplitudes are most efficient to overcome local mechanical constraints without inducing excessive particle rearrangements that will disrupt the formation of dense packings. However, this non-monotonicity progressively weakens with increasing $\mu$, and consequently $\Gamma_{\rm opt}$ is no longer identifiable.  

It is evident that the non-monotonicity of $\phi_{\rm f}(\Gamma)$ is related to the out-of-equilibrium nature of the compaction process: At low $\Gamma$ (typically $\lesssim \Gamma_{\rm opt}$), $\phi(t_w)$ does not fully saturate within the observation time window, indicating the persistence of slow, aging-like dynamics (open symbols in Fig.~\ref{Figure1}(f)). In contrast, for larger $\Gamma$, the system rapidly reaches a steady-state and $\phi$ shows no aging (solid symbols).
This crossover between slow aging and steady dynamics is reminiscent of the dynamic transition observed in thermal glass-formers, where the transformation from an equilibrium supercooled liquid to a nonequilibrium glass is marked by the onset of aging. Analogously, in the granular system, within the glassy regime, $\Gamma \leq \Gamma_{\rm opt}$, compaction proceeds more efficiently, and aging becomes more pronounced as $\Gamma$ approaches the transition threshold at $\Gamma_{\rm opt}$.

At sufficiently large $\Gamma$, the system remains in its initial loose-packed state, i.e.,  $\phi_{\rm f} \cong \phi_0(\mu)$, without further time evolution. We define as $\Gamma_{\rm nc}$ the threshold beyond which no compaction occurs anymore. In other words, the perturbation from the shearing randomizes the configuration, and the system settles into a new configuration that is statistically equivalent to the initial one. This absence of compaction has also been reported in experimental studies of granular packings under strong tapping~\cite{yuan2021experimental}, where high agitation levels continuously disrupt any achieved densification. $\Gamma_{\rm nc}$, marked by the short vertical lines in Fig.~\ref{Figure1}(f), increases monotonically with $\mu$. This trend suggests that for highly frictional grains, stronger perturbations are required to fully fluidize the system and inhibit compaction, consistent with the enhanced mechanical stability conferred by higher friction.

In addition to the ensemble-averaged results, we analyze individual trajectories of $\phi(t_w)$ for three representative cases, shown in Fig.~S2, which highlight the slow and intermittent nature of the compaction process, particularly at low $\mu$. These trajectories suggest that, at low friction, the mechanical network remains fragile, exhibiting intermittent, irreversible large-scale rearrangements, and hence steps in $\phi(t_w)$, that are reminiscent of the avalanche-like events observed in thermal glasses~\cite{kob2000fluctuations,yanagishima2017common}. 
In contrast, at higher $\mu$ (e.g., $\mu \gtrsim 0.3$), these large-scale discontinuous fluctuations are suppressed. Although the precise mechanism responsible for these fluctuations remains to be clarified, one possible interpretation is that strong friction enables local stress accumulations to be relieved through localized snapping events without triggering system-spanning avalanches. This behavior is consistent with the absence of both avalanche-like events and a well-defined $\Gamma_{\rm opt}$. 
To gain deeper insight into the underlying mechanisms, we now analyze the details of the dynamics.

\begin{figure*}
\centering
\includegraphics[width=16.5cm]{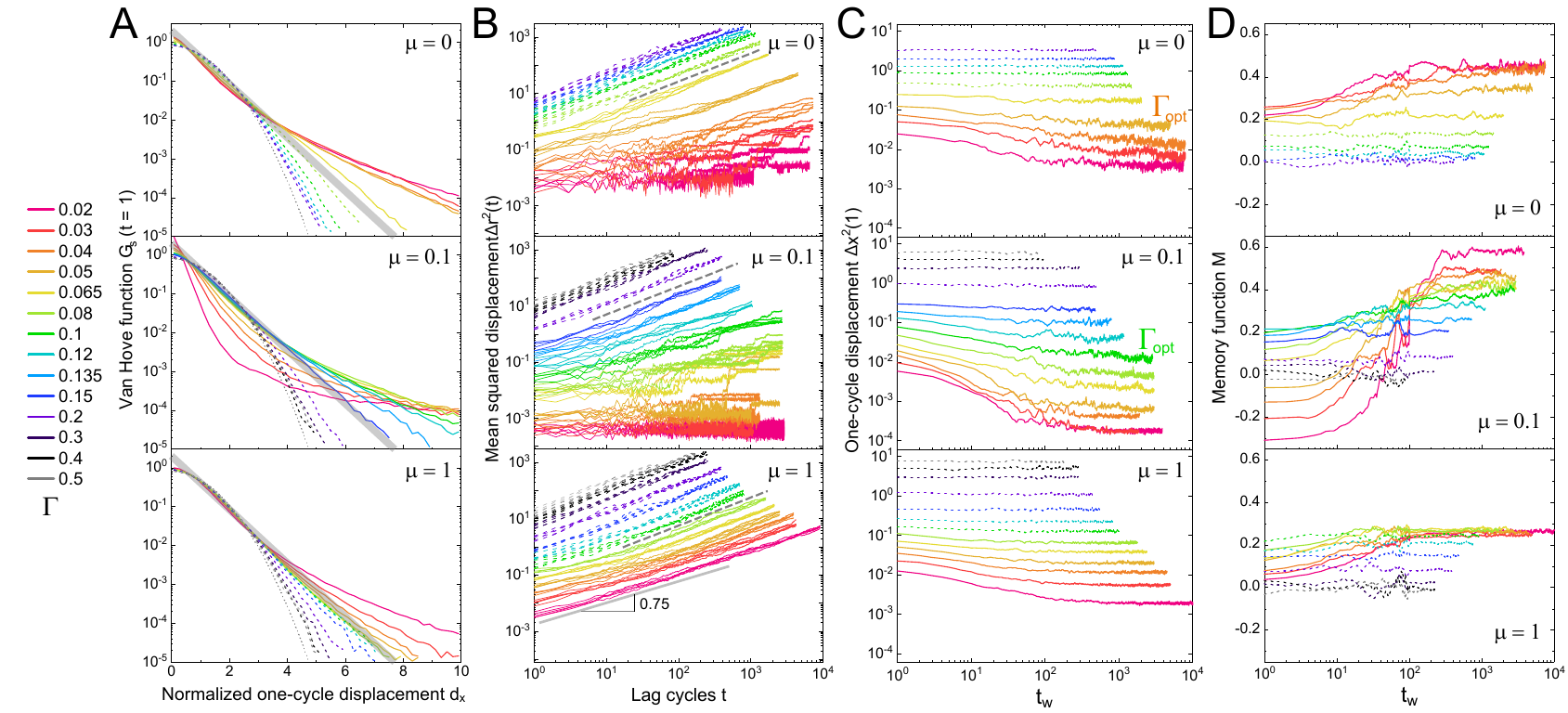}
\caption{
Friction and shear amplitude dependence of aging, memory, and relaxation dynamics. See the color codes in $\Gamma$ on the left. 
(a) Van Hove function $G_s$ at $t=1$, showing the distribution of normalized $x$-directional displacements $d_x = |\delta x_i| / \sqrt{\langle\delta x^2_i\rangle}$ averaged over 300--1000 cycles, excluding the initial transient $t_w\Gamma = 20$. Gray dotted curves represent Gaussian distributions. Bold gray lines represent exponential functions that characterize the decay of $G_s$ in the crossover regime. 
(b) Mean squared displacements $\Delta r^2(t)$ for $t_w\Gamma = 20$. Solid curves indicate the solid-like aging regime with sub-diffusive motion, while dashed curves denote the fluidized regime showing normal diffusion. The solid gray line for $\mu = 1$ marks an intermediate sub-diffusive (creep-like) regime.
(c) One-cycle displacement $\Delta x^2(1) = \langle\delta x_i^2(t_w, 1)\rangle$ for $\mu = 0$, $0.1$, and $1$, smoothed using a filter of size 15 in $t_w$. 
We mark $\Gamma_{\rm opt}$ for $\mu = 0$ and $0.1$.
(d) Memory function corresponding to the data of panel (c), filtered with window sizes of 15 for $t_w \leq 100$ and 50 for $t_w > 100$. All results, except for panel (b), are averaged over 5--10 realizations.
}
\label{Figure2}
\end{figure*}

\subsection*{Characterization of dynamic behaviors}

To quantify the dynamics, we examine the distribution of normalized, one-cycle displacements, i.e., the Van Hove function $G_s(d_x,t)$ at $t=1$, and the mean squared displacement (MSD), $\Delta r^2(t)$, as a function of cycle number, beyond the transient regime ($t_w \geq t^* \approx 20/\Gamma$). These results are shown in Figs.~\ref{Figure2}(a) and (b), with additional data in Fig.~S3. Remarkably, both functions exhibit only a weak dependence on $t_w$ (Fig.~S4), indicating that the properties of the local cage are basically unaffected by the repeated shearing. Furthermore, we confirm that particle displacements exhibit no significant directional dependence (Fig.~S5). 

Figure~\ref{Figure2}(a) shows that with increasing $\Gamma$, $G_s(d_x,t=1)$ evolves from a heavy-tailed, two-step-decay form (solid curves) to a Gaussian-like distribution (dashed curves), reflecting a crossover from heterogeneous, intermittent dynamics to more homogeneous particle motion (see below). At the crossover between these two regimes the tail of the distribution, $d_x \gtrsim 1$, is described well by an exponential decay,
$G_s(d_x,t=1) \sim \exp(-d_x/\xi)$ (thick solid lines), where $\xi\approx 1.58$ denotes the (reduced) characteristic dynamical correlation length. We define the corresponding crossover strain as $\Gamma_{\rm C}$, above which the dynamics become effectively fluid-like, i.e., the system enters a fluidization regime.
In all figures, solid lines indicate data from the solid-like aging regime ($\Gamma < \Gamma_{\rm C}$), while dashed lines denote the fluidized regime ($\Gamma > \Gamma_{\rm C}$).

The MSD data, Fig.~\ref{Figure2}(b), provide complementary insight. They reveal a clear transition at $\Gamma_{\rm C}$ from sub-diffusive to diffusive dynamics with increasing $\Gamma$. (Note that here we are presenting the data for the individual samples.)
For low friction ($\mu = 0$ and 0.1) and small $\Gamma$, the MSDs exhibit strong sample-to-sample fluctuations and pronounced dynamic intermittency. In contrast, for large friction ($\mu = 1$), the MSDs show only weak fluctuations and approach diffusive behavior (dashed lines) for all $\Gamma$, with a sub-diffusive regime at short and intermediate times, which can be associated with creep dynamics~\cite{yuan2024creep} (see solid line; time-and-sample-averaged curves are shown in Fig.~S6). Although the microscopic mechanisms leading to this sub-diffusive dynamics are at present not yet fully understood, their occurrence is consistent with the observed dynamic heterogeneity for $\Gamma < \Gamma_{\rm C}$. Moreover, for moderate $\mu$ and small $\Gamma$, the MSDs display clear caging behavior, indicative of an elastic response~\cite{royer2015precisely,mao2024dynamic}, in line with the plateau observed in $\phi(t_w)$ (Fig.~\ref{Figure1}). Overall, the trends captured by the MSDs are fully consistent with the classification into solid-like and liquid-like regimes based on the Van Hove function. 

To further characterize the nature of the compaction process, we measure the one-cycle mean squared displacement (MSD), $\Delta x^2(t_w, 1)$, and the memory function $M$, which quantifies correlations between displacements for consecutive cycles~\cite{yuan2024creep}, 
see Figs.~\ref{Figure2}(c) and (d).
(Additional data are provided in Fig.~S7). 
Here, the memory function $M$ is defined as:
\begin{equation}
M = -\frac{\langle \delta x(t_w - 1, 1) \delta x(t_w, 1) \rangle}{\langle \delta x^2(t_w, 1)\rangle},
\end{equation}
where $\delta x(t_w, 1)$ denotes the displacement during a single cycle starting at $t_w$. A non-zero $M$ indicates memory in the particle dynamics, with $M > 0$ corresponding to antipersistent motion, where particle displacements tend to be negatively correlated with those in the preceding cycle.

\begin{figure*}
\centering
\includegraphics[width=14cm]{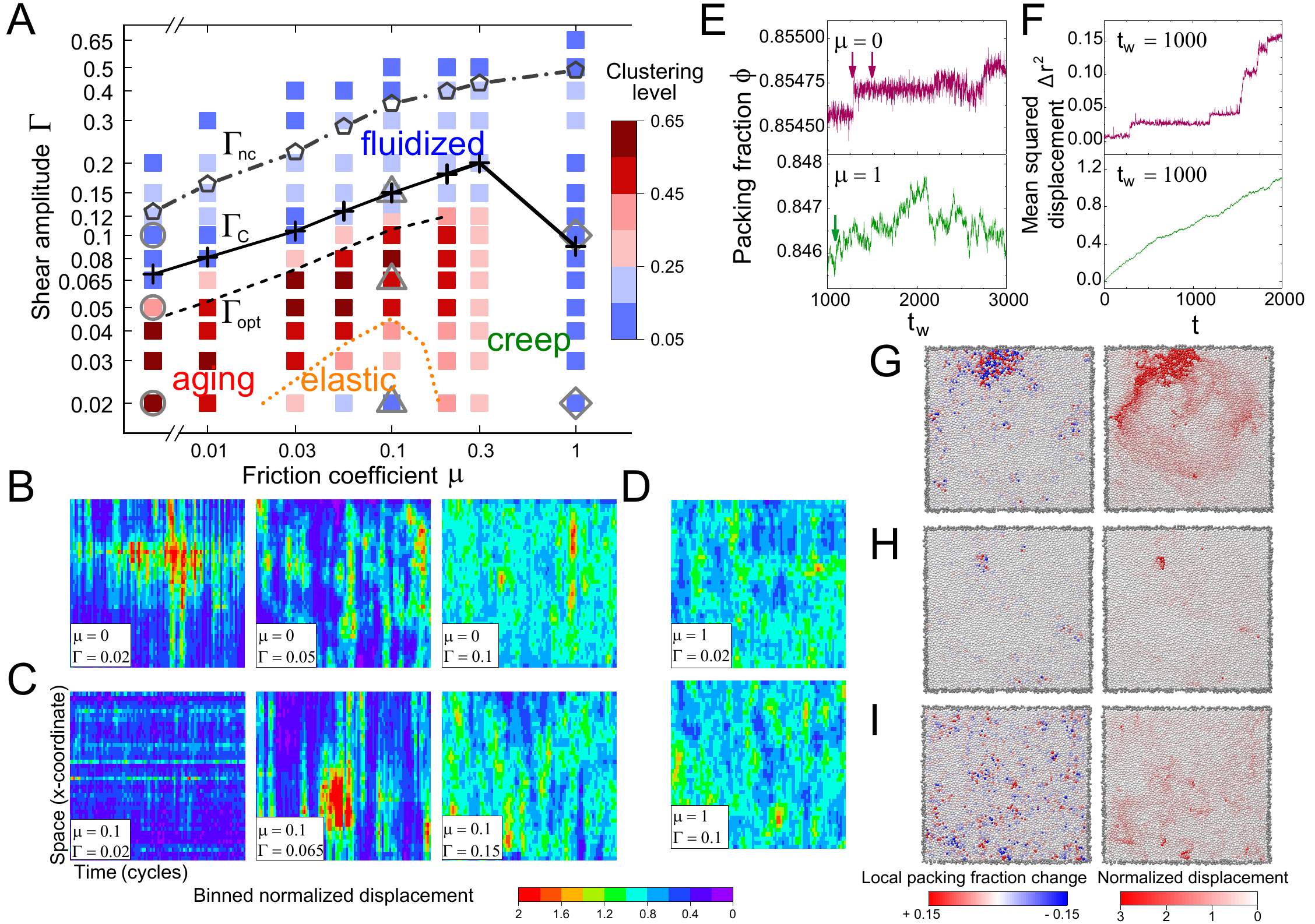}
\caption{
$\Gamma-\mu$ state diagram and dynamic intermittency. See Methods for definitions of local events, normalization, and clustering detection. 
(a) Dynamic state diagram showing the clustering level of the fastest 5\% of particles, averaged over large $t_w$. Clustering level is determined by the ratio of the largest cluster size to the number of selected fast particles. Crosses mark the crossover strain $\Gamma_{\rm C}(\mu)$, identified from the Van Hove function in Fig.~\ref{Figure2}(a). Dashed line shows $\Gamma_{\rm opt}(\mu)$ for $\mu \leq 0.2$ from compaction curves. Dash-dotted line shows the limit strain $\Gamma_{\rm nc}(\mu)$ above which compaction is absent, while the dotted line marks the elastic regime boundary, which is defined via the caged mean squared displacement. 
(b--d) Spatio-temporal maps of normalized particle displacements for selected $\mu$ and $\Gamma$ (marked by gray symbols in (A)). The horizontal axis spans 500 shear cycles, and the vertical axis corresponds to 40 spatial bins along the $x$-direction. 
(e) Packing fraction evolution for $\Gamma = 0.02$ and $\mu = 0$ and 1.0 from $t_w = 1000$ to $3000$. The two arrows indicate the configurations presented in panels (g) and (h). 
(f) Corresponding mean squared displacements $\Delta r^2(t_w = 1000, t)$.
(g)(h) Correlation between local packing fraction changes (left) and normalized particle displacements (right) for $\mu = 0$ at $t_w = 1288$ and $1500$, respectively. $t_w = 1288$ marks a large avalanche event in (e). 
(i) Similar correlations for $\mu = 1$ at $t_w = 1100$. Panels (g-i) share the same color scale (bottom). }
\label{Figure3}
\end{figure*}

$\Delta x^2(1)$ and $M$ display a similar dependence on $t_w$: For $\Gamma < \Gamma_{\rm C}$, $\Delta x^2(1)$ undergoes a rapid initial decrease, and $M$ shows a sharp increase, followed by a more gradual evolution.
In contrast, for $\Gamma > \Gamma_{\rm C}$, the dynamics rapidly stabilize into a steady state with minimal aging and weak memory effects. Interestingly, for $\Gamma \approx \Gamma_{\rm opt}$ and $\mu = 0$ or $0.1$, the system exhibits a prolonged, slow evolution in $\Delta x^2(1)$ as well as in the memory function $M$. The associated $t_w$-dependence can be described well by a power-law with a small exponent or alternatively by a logarithmic law, although the exact $t_w$-dependence remains to be confirmed. This behavior parallels the slow compaction observed in $\phi(t_w)$ (Figs.~\ref{Figure1}(c--e)), further supporting the identification of this regime with mechanical aging. 

In addition, Fig.~S8(a) reveals that, for $\mu \leq 0.1$, the value of $\Delta x^2(1)$ at $\Gamma_C$ is approximately 0.2, indicating identical dynamics in terms of absolute displacements (not only the exponential shape of $G_s$).
This suggests that mild friction enhances mechanical stability in a monotonic manner. However, this no longer holds for larger friction (Fig.~S8(b)), where the nature of the dynamics fundamentally changes to a creep-like, continuous form. In Figs.~S8(d--h), the parametric plot of $\Delta x^2(1)$ versus $M$ yields a master curve in the high-$\Gamma$ fluidized regime, while clear aging signatures emerge at low $\Gamma$. 

From these results, it is clear that the dynamic crossover at $\Gamma_{\rm C}$ in Figs.~\ref{Figure2}(a)(b) corresponds to the transition between slow aging and steady state, as illustrated in Figs.~\ref{Figure1}(f),~\ref{Figure2}(c)(d) (see the distinction between solid and dashed curves). 

\subsection*{Dynamic state diagram}

Based on these dynamic characteristics, we construct a state diagram as a function of $\Gamma$ and $\mu$, Fig.~\ref{Figure3}(a), which exposes how friction and shear amplitude govern structural evolution and dynamic transitions in granular compaction. Key characteristic strain amplitudes include: The upper limit of the elastic regime, $\Gamma_{\rm ela}$; the optimal compaction amplitude, $\Gamma_{\rm opt}$; the onset of fluidization, $\Gamma_{\rm C}$; and the threshold for complete randomization, $\Gamma_{\rm nc}$.

To further characterize the dynamics in the aging regime, we measure the clustering level of fast particles across all systems (represented by colored symbols in Fig.~\ref{Figure3}(a)), which informs on the degree of dynamic intermittency (see Methods for details).  For selected state points (gray open symbols in Fig.~\ref{Figure3}(a)), Figs.~\ref{Figure3}(b--d) display the spatio-temporal patterns of particle displacements. These reveal starkly different patterns: Strongly intermittent rearrangements at low friction and $\Gamma < \Gamma_{\rm C}$ (panels (b) and (c)), and continuous, mildly heterogeneous dynamics at high friction (panel (d)).

Figures~\ref{Figure3}(e) and (f) compare $\phi(t_w)$ and the associated MSDs from individual trajectories for $\Gamma = 0.02 < \Gamma_{\rm C}$ and $\mu = 0$ and $1$. For $\mu = 0$, the system exhibits strongly intermittent dynamics, with large, correlated jumps in both $\phi(t_w)$ and the MSD. Visualization of local packing fraction changes and displacement fields (Figs.~\ref{Figure3}(g) and (h)) confirms that fast particle motion coincides with local volume fluctuations, indicative of avalanche-like rearrangements, as also observed in deeply quenched thermal glasses~\cite{kob2000fluctuations,yanagishima2017common}. In contrast, for $\mu = 1$, the correlation between fluctuations in $\phi(t_w)$ and the MSD is weak, and the displacement field exhibits only moderate heterogeneity, Fig.~\ref{Figure3}(i). Figure~S9 further supports that, for $\Gamma < \Gamma_{\rm C}$, compaction is generally spatially correlated with collective dynamics, though this correlation is gradually suppressed with increasing $\mu$.

Unlike thermal model glasses, where cyclic shear triggers a sharp yielding transition~\cite{leishangthem2017yielding,yeh2020glass}, granular systems near jamming exhibit complex relaxation and a more gradual yielding process~\cite{mailman2014consequences,dagois2017softening}. Notably, the heterogeneous sub-diffusion, intermediate between caged and diffusive dynamics, gives rise to pronounced mechanical aging, as it enables the slow optimization of the packing structure.

Furthermore, friction governs the microscopic pathways of the aging process in configuration space. At small $\mu$, the density of mechanically stable states in configurational space is rather low, leading to intermittent, avalanche-like dynamics and a stepwise evolution towards denser states. As $\mu$ increases, the number of accessible stable states grows quickly, enabling more frequent but smaller rearrangements. The nature of aging thus shifts from intermittent to creep-like, governed by the interplay between frictional stabilization and the system’s ability to explore configuration space. To clarify the reentrant friction dependence observed in the aging regime, we next analyze the full-cycle shear response, moving beyond the stroboscopic measurements discussed above.

\begin{figure}[!t]
\centering
\includegraphics[width=8cm]{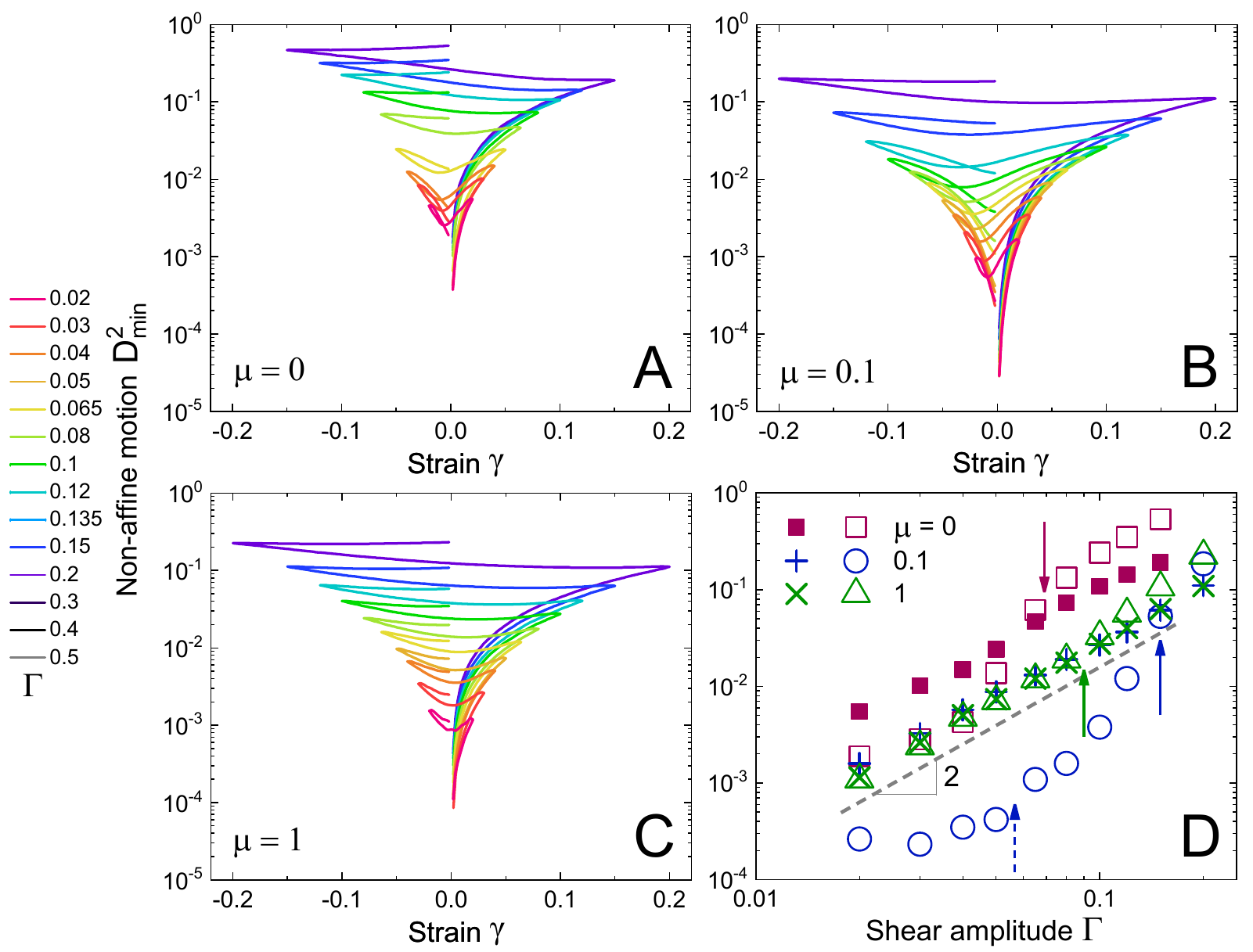}
\caption{(a--c) Average non-affine motion $D^2_{\rm min}$ over a complete strain loop ($\gamma = 0 \rightarrow \Gamma \rightarrow -\Gamma \rightarrow 0^*$) for three friction coefficients $\mu$ and various shear
 amplitudes $\Gamma$. See the color code for $\Gamma$ on the left.
 (d) Comparison of $D^2_{\rm min}$ at maximal strain $\gamma = \Gamma$ ($\blacksquare$, +, x) and after strain reversal $\gamma = 0^*$ ($\square$, $\bigcirc$, $\triangle$) for $\mu=0, 0.1$, and 1, respectively. The dashed line shows  the scaling relation of $\Gamma^2$. Arrows mark the crossover strains $\Gamma_{\rm C} = 0.045$ for $\mu=0$ and $\Gamma_{\rm C}= 0.11$ for $\mu=0.1$, above which irreversible motion becomes dominant, as well as the elastic-aging boundary $\Gamma \approx 0.05$ for $\mu = 0.1$.} 
\label{Figure4}
\end{figure}

\subsection*{Elastoplasticity in single shear cycle}

To investigate the late-stage dynamics, defined as the final $20\%$ of cycles in the compaction process, as a function of $\mu$ and $\Gamma$, we track the averaged non-affine motion $D^2_{\rm min}$ over a complete shear cycle (strain loop: $\gamma = 0 \rightarrow \Gamma \rightarrow -\Gamma \rightarrow 0^*$).  We define $D^2_{\rm min} = \langle \vec{d}_i^2(\gamma) \rangle$,
where $\langle.\rangle$ denotes the average over all particles, and the non-affine displacement of particle $i$ is given by $\vec{d}_i(\gamma) = \delta \vec{r}_i(\gamma) - \Sigma_j \delta \vec{r}_j(\gamma)/N_i$, with the sum taken over all contacting neighbors of $i$. Here, $\gamma = 0$ serves as the reference state within the shear cycle. This contact-based definition reflects the mechanically relevant local environment in frictional granular systems. 

Figures~\ref{Figure4}(a--c) show that, for small $\Gamma$ and basically independent of $\mu$, $D^2_{\rm min}$ exhibits local maxima at $\gamma = \pm \Gamma$ and remains essentially reversible. This indicates that after deformation under one-sided simple shear, the system partially recovers its elastic response. As $\Gamma$ increases, reversibility diminishes and a pronounced zig-zag pattern emerges, reflecting the increasing dominance of irreversible (plastic) motion. This evolution is quantified in Fig.~\ref{Figure4}(d), where we compare $D^2_{\rm min}(\gamma = 0^*)$ with $D^2_{\rm min}(\gamma = \pm \Gamma)$. For small $\mu$, the crossover between these two motions aligns closely with $\Gamma_{\rm C}$, extending the concept of a yielding point previously observed in soft glasses~\cite{keim2014mechanical, galloway2022relationships}. In contrast, for $\mu = 1$, reversibility is weak for all $\Gamma$, and the transition near $\Gamma_{\rm C}$ becomes more gradual. These behaviors are further illustrated by the spatially resolved $\vec{d}_i^2(\gamma)$ fields shown in Fig.~S10.

Figure~\ref{Figure4}(d) shows that $D^2_{\rm min}(\gamma = \Gamma)$ scales approximately as $\Gamma^2$ for all $\mu$, but with a prefactor that is largest for $\mu=0$. This confirms the stabilizing role of frictional contacts, in that the deformation proceeds through simpler, more localized rearrangements. This trend is also consistent with the increase of the compaction threshold $\Gamma_{\rm nc}(\mu)$ with $\mu$ (dash-dotted line in Fig.~\ref{Figure3}(a)).
As $\mu$ approaches 1, reversibility is strongly suppressed upon strain reversal, as can be deduced from the weak recovery in $D^2_{\rm min}$ and reduced memory effects (Fig.~\ref{Figure2}(d)). This reflects another effect of friction, apart from changing the dynamics from intermittent to creep-like as mentioned in previous section. The resulting reentrant trend with increasing $\mu$ arises from a competition between friction-induced stabilization and the loss of reversibility. This delicate balance, shaped by frictional contact constraints~\cite{mao2024dynamic}, governs the system's  relaxation pathways and overall dynamic response under cyclic shear. Notably, for $\mu = 0.1$, the existence of an elastic regime at low $\Gamma$ leads to a qualitatively distinct post-deformation behavior (dashed arrow in Fig.~\ref{Figure4}(d)) compared to the case $\mu =0$ and $1$.

\section*{Discussion}

Our study reveals the dual role of friction: It can stabilize granular assemblies, but  also fluidize them, leading to a reentrant transition between aging and flow. 
This behavior is demonstrated by numerically exploring the compaction dynamics of disordered granular matter under cyclic shear across a range of strain amplitudes ($\Gamma$) and friction coefficients ($\mu$). 
Over many cycles, the system’s response --- characterized by the packing fraction ($\phi$) and particle dynamics --- exhibits three distinct regimes: elastic, aging, and fluidized (Fig.~\ref{Figure3}(a)). 
The aging regime, marked by logarithmic-like slow compaction, emerges between the elastic and fluidized states, in contrast to the sharp elasto-plastic yielding transitions typically observed in frictionless amorphous solids.

Thus, the long-standing discrepancies between logarithmic-like and steady-state compaction behaviors in previous experiments of granular compaction~\cite{richard2005slow} can be rationalized by the presence of two distinct regimes --- aging and fluidized --- separated by a crossover driving strength $\Gamma_{\rm C}$. Friction strongly reshapes the system's response:  As $\mu$ increases, the number of accessible stable configurations grows, enhancing mechanical stability in the forward shear direction while reducing reversibility during shear reversal at high friction. Consequently, the dynamics evolves from intermittent to creep-like at small $\Gamma$. Thus, friction emerges as a two-edged control parameter: It suppresses avalanches by stabilizing contacts, yet ultimately accelerates fluidization by opening continuous relaxation pathways, giving rise to a reentrant slowdown at intermediate friction~\cite{royer2015precisely,mao2024dynamic}.

Previous studies have reported disparate dynamical regimes under cyclic shear --- caging in idealized soft-sphere~\cite{kawasaki2016macroscopic,das2020unified} and glass models~\cite{leishangthem2017yielding,yeh2020glass} or sub-diffusion in marginally jammed (frictionless) granular systems~\cite{mailman2014consequences,dagois2017softening} for small $\Gamma$, and long-term diffusion in highly frictional grains for all $\Gamma$~\cite{royer2015precisely,mao2024dynamic,yuan2024creep}. 
Our state diagram unifies these behaviors by revealing how the interplay between $\mu$ and $\Gamma$ controls the aging dynamics (Fig.~\ref{Figure2}(b); Figs.~S3(f--j)), showing that sub-diffusion at small $\Gamma$ consistently corresponds to dynamical heterogeneity and aging. This robustness merits further investigation in the broader context of anomalous diffusion~\cite{metzler2014anomalous}, where geometric heterogeneity and constraint/memory in particle motion may play distinct roles.

Unlike aging monitored through particle-level rearrangements, contact-level aging has been observed experimentally in granular matter~\cite{murphy2020memory,farain2024perturbation,farain2024thermal}. In contrast to our controlled finite strain amplitudes ($\Gamma$), real granular materials relax through complex, multi-scale perturbations, including persistent creeping even in the absence of clear mechanical drives~\cite{deshpande2021perpetual}. Although our model does not capture these intricate microscopic processes, it elucidates the essential roles of jamming marginality and friction, highlighting subtle yet crucial deviations from standard glassy dynamics. The complex relaxation behaviors revealed here carry broad implications --- from optimizing powder compaction and industrial granular flows to interpreting slow deformation in soils and seismic fault dynamics, where friction and mechanical noise govern large-scale material responses.

Although our simulations are restricted to two-dimensional disk packings, the underlying mechanism --- frictional control of metastable states and reentrant aging--fluidization --- is geometric and thus expected to extend also to three dimensions. Future cyclic shear experiments on 3D packings with tunable surface friction (e.g. by varying roughness or coatings) could directly test this prediction.

Taken together, our results establish a coherent framework that unifies granular compaction, creep, and fluidization as manifestations of the same friction-controlled nonequilibrium dynamics. Rather than emphasizing how friction stabilizes or fluidizes, this perspective highlights its role in governing the accessibility of metastable states and the routes by which jammed assemblies relax. This principle extends beyond granular materials, offering a general view of how microscopic constraints dictate macroscopic aging and flow in athermal disordered systems, including frictional colloids and biological assemblies.

\section*{Methods}

\subsection*{Simulation details} 

We perform Discrete Element Method (DEM) simulations to model two-dimensional granular materials composed of frictional disks. Particle interactions are governed by a linear spring--dashpot model incorporating static friction~\cite{silbert2001granular}. To suppress crystallization and ensure a disordered structure, we employ a bidisperse mixture with a 50:50 composition and a size ratio of 1.4, with 5000 particles in total. For two particles in contact, the normal and tangential forces acting on particle $i$ are given by: 
\begin{equation}
\vec{F_n} = k_n\delta_{ij}\vec{n}_{ij} - \gamma_nm_{\rm eff}\vec{v}_n,
\end{equation}
\begin{equation}
\vec{F_t} = -k_t\vec{u}_t - \gamma_tm_{\rm eff}\vec{v}_t,
\end{equation}
where $k_n$ and $k_t$ are the spring constants (here we set $k_t = 2/7k_n$), $\gamma_n$ and $\gamma_t$ are the damping coefficients, $\delta_{ij}$ and $\vec{n}_{ij}$ denote, respectively, the overlap distance and the unit vector pointing from particle $i$ to particle $j$, respectively, $\vec{v}_n$ and $\vec{v}_t$ are the relative normal/tangential velocities, $m_{\rm eff} = \frac{m_i m_j}{m_i + m_j}$ is the effective mass, and $u_t = \int v_t dt$ is the sliding distance during contact. The restitution coefficients $e_{n,t}$ and the collision time $t_{\mathrm{col}}$ are related by $e_{n,t} = \exp(-\gamma_{n,t} t_{\rm col} / 2)$, with $t_{\rm col} = \pi/\sqrt{2k_n / m - \gamma_n^2 / 4}$ (with $m$ denoting the mass of a small disk). The tangential force magnitude $|\vec{F_t}|$ is truncated at $\mu|\vec{F_n}|$ when $|\vec{F_t}| > \mu|\vec{F_n}|$, where $\mu$ denotes the static friction coefficient. We set $e_{n,t} = 0.5$ and the simulation time step as $0.012 t_{\rm col}$. The radius of small disks is defined as unit length ($a = 1$), and the areal particle density is $\rho = 1$. 

For each value of $\mu$, we generate isotropically jammed packings under periodic boundary conditions, applying a standard quasistatic compression protocol~\cite{o2002random, silbert2010jamming, yuan2021universality}. Starting from a dilute, overlap-free configuration with packing fraction $\phi < 0.1$, the system undergoes successive compression--relaxation steps. In each step, the packing fraction is increased by a small increment, $\Delta \phi = 2 \times 10^{-4}$, followed by relaxation using a DEM procedure, with the volume held fixed. The relaxation continues until the unbalanced stress is fully released, guaranteeing that the system remains unjammed before the next compression step is applied. After several such iterations, the system reaches a mechanically stable configuration and becomes jammed at a packing fraction $\phi_0$. The chosen increment $\Delta \phi$ is sufficiently small to avoid that the final jamming density $\phi_0(\mu)$ is significantly influenced.

Particles located within a distance of $4a$ from the packing cell boundary are designated as fixed boundary layers, as illustrated in Fig.~\ref{Figure1}(b). The two layers along the $x$-direction rotate about their centers to induce shear, while the two layers along the $y$-direction deform adaptively in response to the $x$-directional motion, shifting in the $y$-direction to maintain a constant target pressure $P$. The pressure $P$ is computed as the trace of the Cauchy stress tensor, based on interparticle contact forces. This constant-pressure shear geometry allows the packing to accommodate volume changes, such as compaction and dilation, during deformation.

Once the shear protocol is initiated, the dimensionless pressure $\tilde{P} = P/(k_n a^2)$, which approximates the typical particle overlap, is fixed at $\tilde{P} = 4 \times 10^{-3}$ to ensure that the system remains close to the hard-particle jamming point. This pressure corresponds to an excess packing fraction of approximately 0.01 beyond the jamming onset. To maintain quasistatic conditions, a small shear rate $\dot{\gamma}$ is applied such that the inertia number $I = \dot{\gamma} a / \sqrt{P/\rho}$ is kept at $2 \times 10^{-4}$, placing the system deep within the quasistatic regime.

The shear cycle follows the deformation protocol: $\gamma = 0 \rightarrow \Gamma \rightarrow 0 \rightarrow -\Gamma \rightarrow 0$. The main analysis focuses on the stroboscopic configurations recorded after each full shear cycle, indexed by $t$. Starting from a loosely jammed packing, the system is subjected to cyclic shear until the accumulated strain reaches $4 \Gamma t \approx 400$--$800$, depending on the value of $\Gamma$. For each combination of $\mu$ and $\Gamma$, we perform between 4 and 10 independent simulations, and report the corresponding statistical averages. In Fig. 4, we further analyze ten complete shear cycles with high resolution $\delta\gamma = 0.002$ for selected friction values $\mu = 0$, $0.1$, and $1$.

\subsection*{Intermittency analysis}

For an individual particle trajectory, we define a coarse-grained displacement $\Delta x_i(t) = (\sum_{t'=t - t_f + 1}^{t} x_i(t') - \sum_{t'=t+1}^{t + t_f} x_i(t'))/t_f$ which suppresses mechanical noise and helps identify local rearrangement events, following a method adapted from~\cite{candelier2009building, schoenholz2016structural}. We choose a filter time of $t_f = 15$, which provides a good balance between noise reduction and temporal resolution; varying $t_f$ does not significantly affect the subsequent analyses. Analogously, we define $\Delta y_i(t)$ and the two-dimensional displacement magnitude $\Delta r_i^2(t) = \Delta x_i^2(t) + \Delta y_i^2(t)$. This approach is also applied to measure local packing fraction, defined for each particle $i$ as the ratio of its area $A_i$ to the area of its Voronoi cell $V_i$: $\phi_i(t)=A_i/V_i(t)$. The coarse-grained change in local packing, denoted $\Delta \phi_i(t)$, captures local dilation or compaction (see left panels of Figs.~\ref{Figure3}(g--i)).

Instead of analyzing the distribution of $\Delta r_i^2(t)$ at a single time point, we evaluate it over a finite time window $t \in [t_w, t_w + \Delta t]$, with $\Delta t = 100 - 500$. Within this interval, each value of $\Delta r_i^2(t)$ is normalized by the spatio-temporal average computed across all particles and time steps within the same window (see right panels of Figs.~\ref{Figure3}(g--i)). This normalization accounts for variations in the overall activity level across different conditions, enabling a direct comparison of spatial heterogeneity patterns while factoring out global aging effects.

We identify the top 5\% of fast particles based on their displacement magnitude $\Delta r_i^2(t)$ within a given time window $t\in[t_w, t_w + \Delta t]$. We then track their spatial clustering over time. Two labeled particles are considered part of the same cluster if they are in contact. While this criterion is more restrictive than using nearest-neighbor distance, it provides a robust and parameter-free way to capture mechanically relevant clusters. The clustering level is defined at each time step as the ratio of the size of the largest cluster to the total number of labeled particles (which is constant), and then averaged over the full time window. The results are insensitive to the exact choice of fast-particle threshold, provided both displacements and cluster sizes are normalized consistently. This metric exhibits only weak dependence on $t_w$, aside from the initial transient state (see Fig.~S4(d)), and is used to construct the state diagram in Fig.~\ref{Figure3}(a) after averaging over late-stage $t_w$. 

To visualize spatio-temporal intermittency, we normalize $\Delta r_i^2(t)$ and divide the particle $x$-coordinates into 40 spatial bins of equal width. At each time step, we compute the bin-averaged normalized displacement and display the result as a heatmap, using the physical $x$-coordinate along the spatial axis and time along the temporal axis, as shown in Figs.~\ref{Figure3}(b--d).

\section*{Acknowledgments}

H.T. acknowledges the support by the Grant-in-Aid for Specially Promoted Research (JSPS KAKENHI Grant No. JP20H05619) from the Japan Society for the Promotion of Science (JSPS). W.K. is a senior member of the Insitut Universitaire de France.



\providecommand{\noopsort}[1]{}\providecommand{\singleletter}[1]{#1}%

\setcounter{figure}{0}
\renewcommand{\thefigure}{S\arabic{figure}}

\begin{figure*}
\centering
\includegraphics[width=16cm]{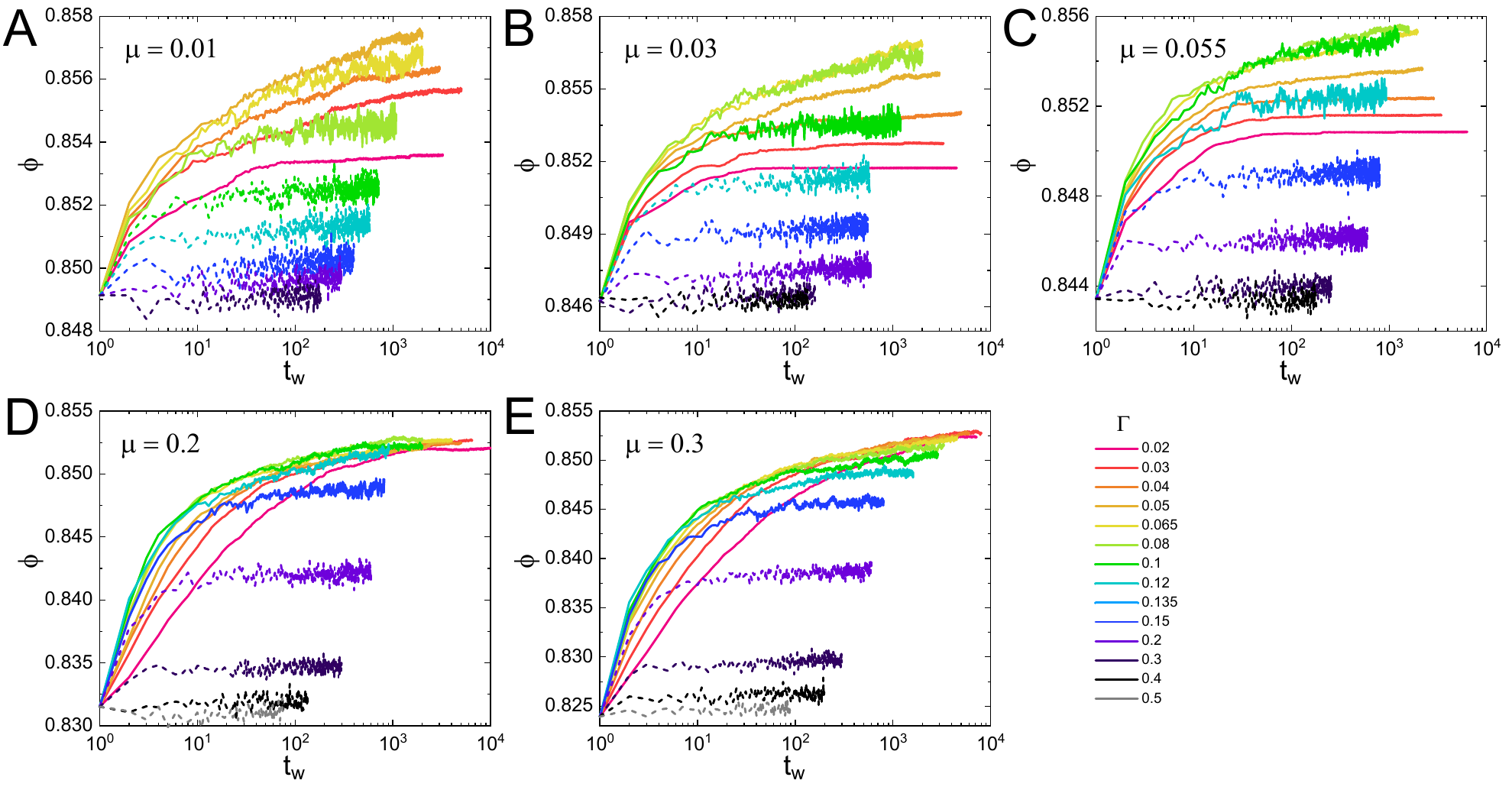}
\caption{Time evolution of packing fraction for varying friction coefficients and strain amplitudes. Temporal evolution of the packing fraction, $\phi(t_w)$, for five representative friction coefficients, $\mu$, under various strain amplitudes, $\Gamma$ (color-coded as indicated in the bottom right). 
Complementing Figs.~1(c--e) in the main text, the optimal strain amplitude for compaction, $\Gamma_{\rm opt}$, increases with $\mu$ for $\mu \lesssim 0.2$, but becomes less well-defined at higher $\mu$. In contrast, the threshold strain above which compaction is no longer effective, $\Gamma_{\rm nc}$, increases monotonically with $\mu$.}
\label{FigureS1}
\end{figure*}

\begin{figure*}
\centering
\includegraphics[width=16cm]{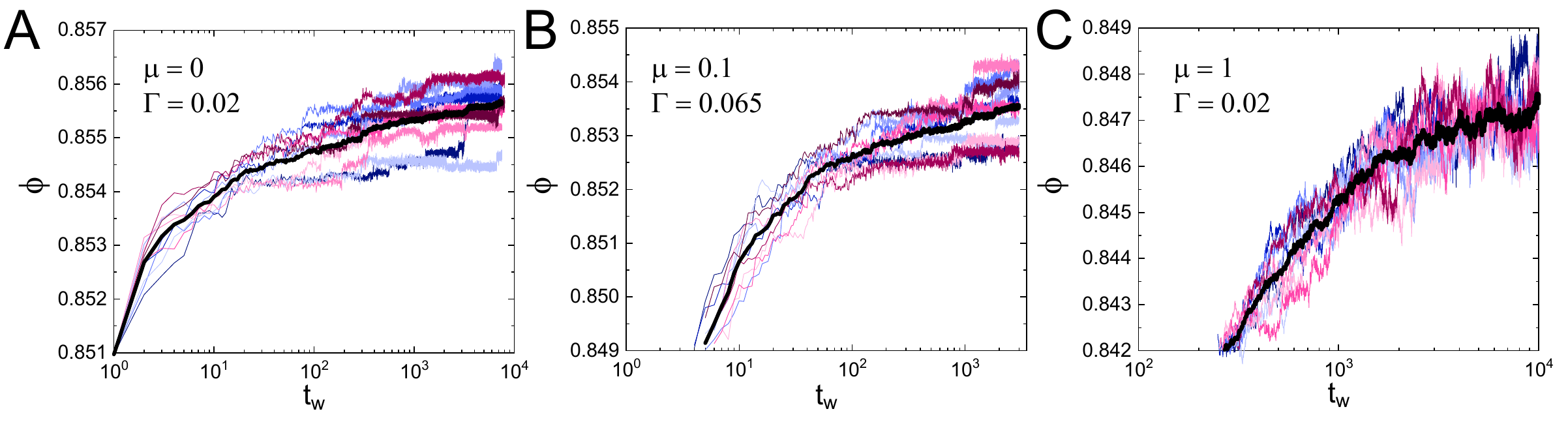}
\caption{Compaction dynamics for multiple realizations. Packing fraction evolution, $\phi(t_w)$, for three systems, each comprising ten independent realizations. Thick black curves indicate ensemble averages. Panels (b) and (c) show magnified views of the corresponding $\phi$ ranges in (a), allowing clearer visualization of the compaction behavior.}
\label{FigureS2}
\end{figure*}

\begin{figure*}
\centering
\includegraphics[width=16cm]{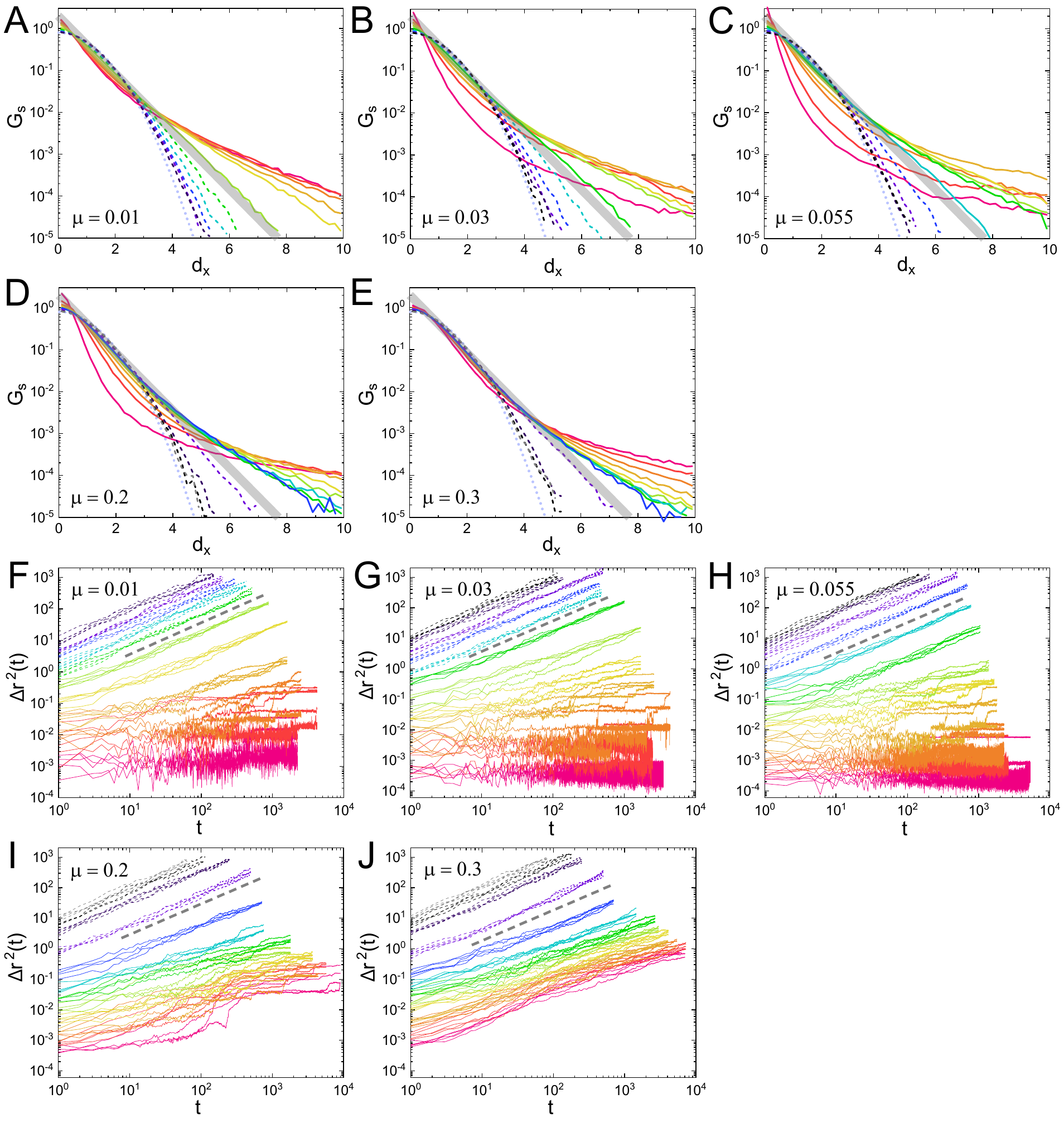}
\caption{Van Hove function and mean squared displacement. Color code is the same as in Fig.~S1.
(a--e) Van Hove function $G_s$ of one-cycle normalized $x$-directional displacements for five different friction coefficients $\mu$, excluding the initial transient.
Dotted curves represent Gaussian fits, while thick gray lines indicate exponential decay, illustrating the emergence of heavy-tailed distributions.
(f--j) Mean squared displacement $\Delta r^2(t)$, computed beyond the initial transient for $t_w \Gamma \approx 20$.
Dashed lines denote the expected scaling for normal diffusion.
}
\label{FigureS3}
\end{figure*}

\begin{figure*}
\centering
\includegraphics[width=15cm]{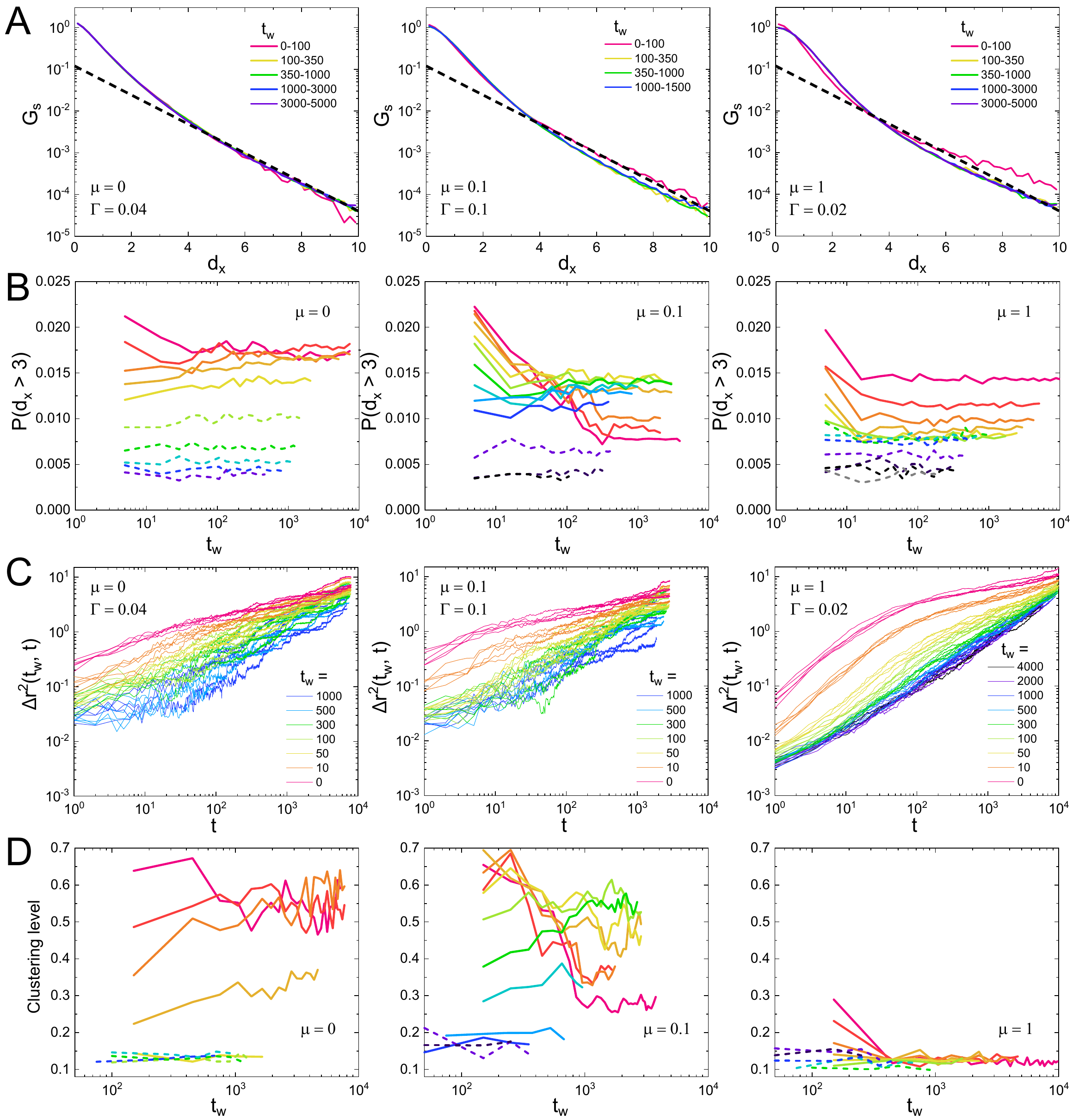}
\caption{Displacement dynamics and clustering behavior.
(a) Van Hove function $G_s$ of one-cycle displacement for three selected systems.
Different shear cycle intervals are used to compute $G_s$. Black dashed lines indicate, as reference, an exponential decay, showing the mild change in the tails for different $\mu$. 
(b) Probability of large displacements in $G_s$, quantified by $P(d_x > 3)$, plotted as a function of waiting time ($t_w$) for three values of $\mu$ and various shear amplitudes $\Gamma$ (color-coded as in Fig.~S1). This confirms that for all $\Gamma$ the shape of $G_s$ depends only weakly on $t_w$, once the transient is properly excluded.
(c) Mean squared displacement $\Delta r^2(t_w, t)$ for the same systems shown in (a).
Intermittent dynamics persist for $\mu = 0$ and $\mu = 0.1$.
(d) Clustering level of fast particles (see Methods for definition) as a function of $t_w$, using the same color code as in Fig.~S1.
The weak dependence on $t_w$, aside from the initial transient, supports the robustness of the liquid--solid crossover strain $\Gamma_{\rm C}$ --- particularly for $\mu \lesssim 0.2$ --- as shown in Fig.~3(a), irrespective of aging or compaction effects.
}
\label{FigureS4}
\end{figure*}

\begin{figure*}
\centering
\includegraphics[width=16cm]{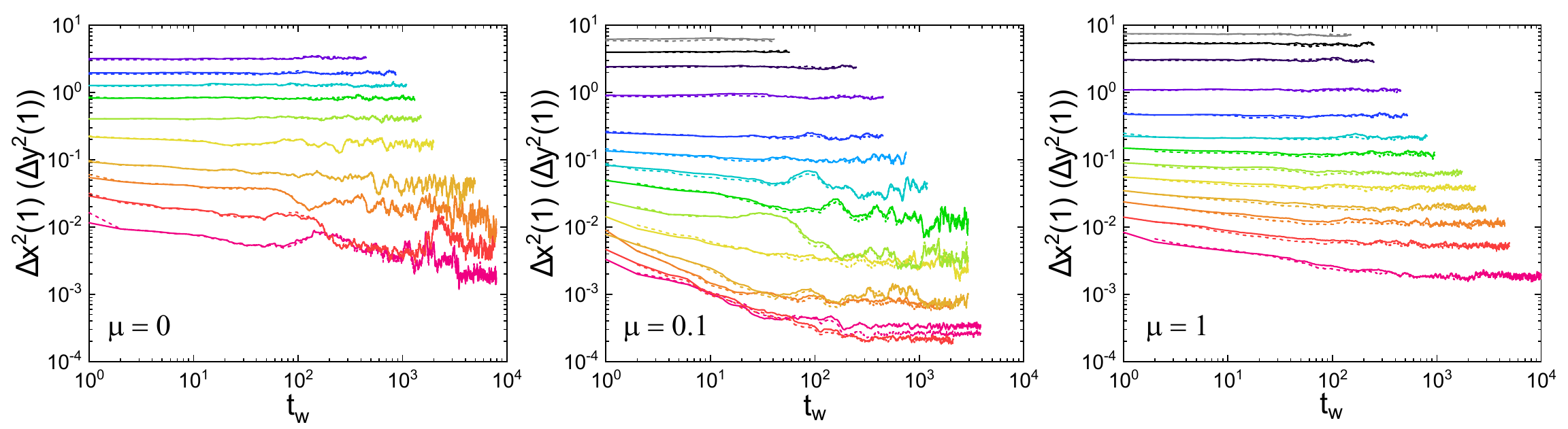}
\caption{
One-cycle displacements in the $x$-direction ($\Delta x^2(1)$, solid lines) and $y$-direction ($\Delta y^2(1)$, dashed lines) as functions of waiting time ($t_w$), shown for three different friction coefficients ($\mu$) and a range of shear amplitudes ($\Gamma$). Each curve corresponds to a single representative simulation trajectory per parameter set.
}
\label{FigureS5}
\end{figure*}

\begin{figure*}
\centering
\includegraphics[width=16cm]{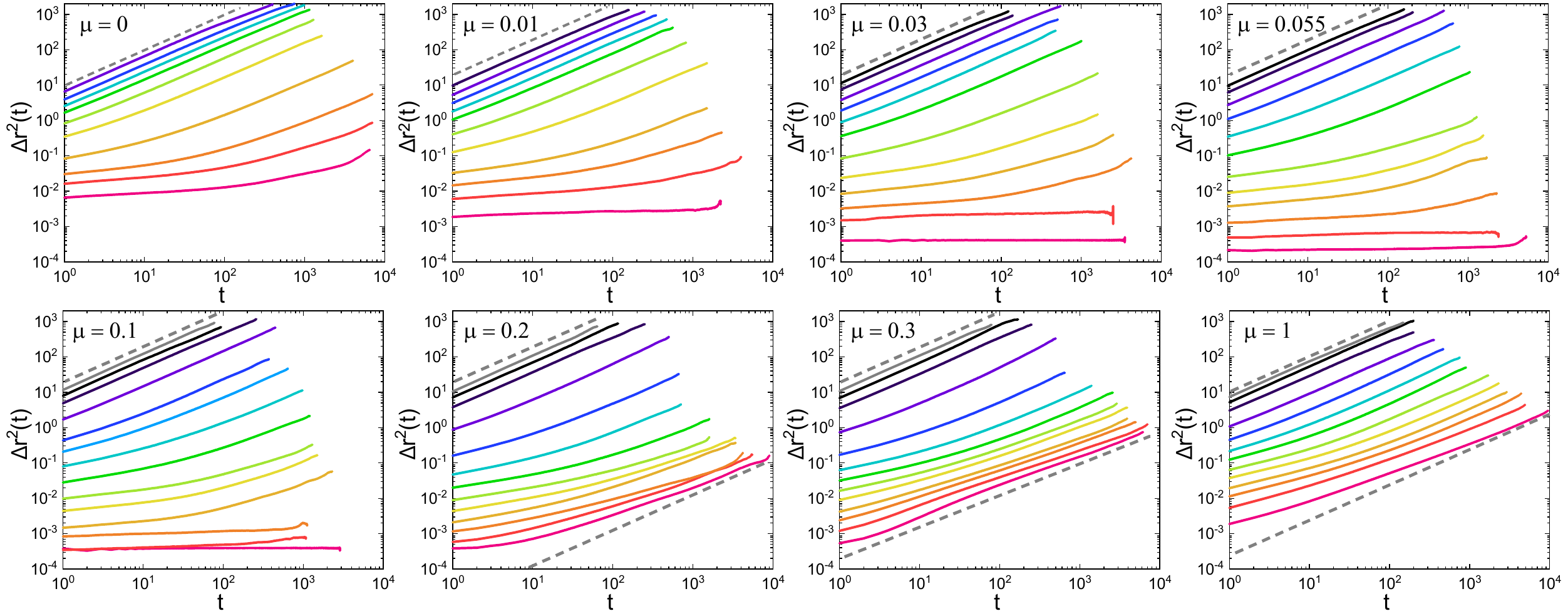}
\caption{Time-and-sample-averaged MSDs beyond the initial transient ($t_w \Gamma \approx 20$). Sub-diffusion is clearly observed for small $\Gamma$. Dashed lines denote the normal diffusion.
}
\label{FigureS6}
\end{figure*}

\begin{figure*}
\centering
\includegraphics[width=16cm]{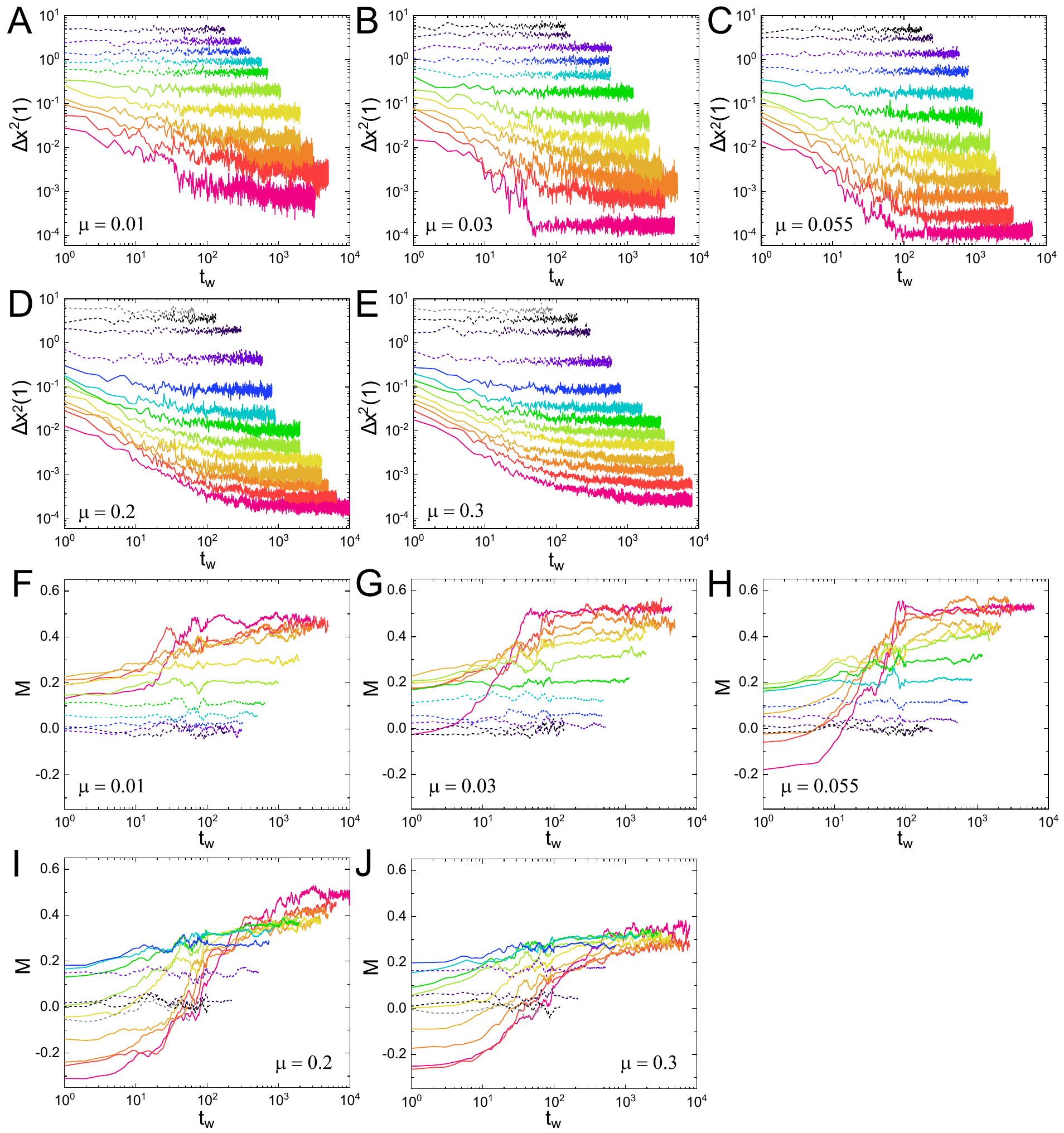}
\caption{One-cycle displacement and memory dynamics for different friction coefficients.
(a--e): One-cycle displacement $\Delta x^2(1)$; (f--j): Memory function $M$, both plotted as functions of waiting time ($t_w$) for the same systems shown in Fig.~S1 (color-coded accordingly). To suppress noise in the memory data, a smoothing filter of width 15 is applied for $t_w \leq 100$, and a width of 50 for $t_w > 100$.}
\label{FigureS7}
\end{figure*}

\begin{figure*}
\centering
\includegraphics[width=15cm]{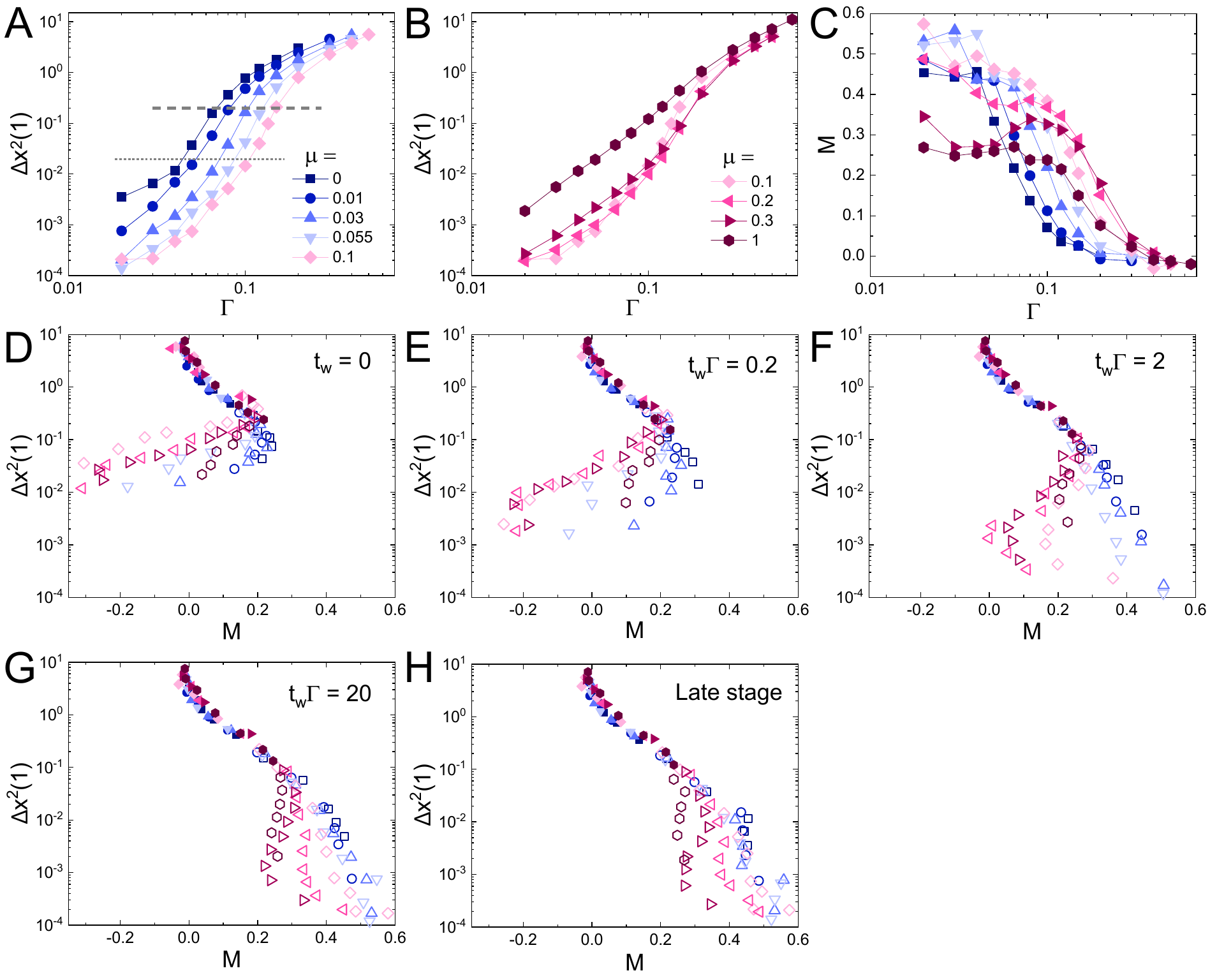}
\caption{Late-stage displacement and memory dynamics.
(a)(b) Late-stage one-cycle displacement $\Delta x^2(1)$ as a function of $\Gamma$ for all values of $\mu$.
Dotted and dashed lines in panel a (for small $\mu$) indicate constant $\Delta x^2(1)$ at $\Gamma_{\rm opt}(\mu)$ and $\Gamma_{\rm C}(\mu)$, respectively. 
(c) Late-stage memory $M$ for all systems.
(d--h) Scatter plots of $\Delta x^2(1)$ versus $M$ for all systems at various values of $t_w \Gamma$.
For each $\mu$ (indicated by color), filled symbols represent data points with $\Gamma > \Gamma_{\rm C}$, while open symbols indicate $\Gamma < \Gamma_{\rm C}$.
}
\label{FigureS8}
\end{figure*}

\begin{figure*}
\centering
\includegraphics[width=9cm]{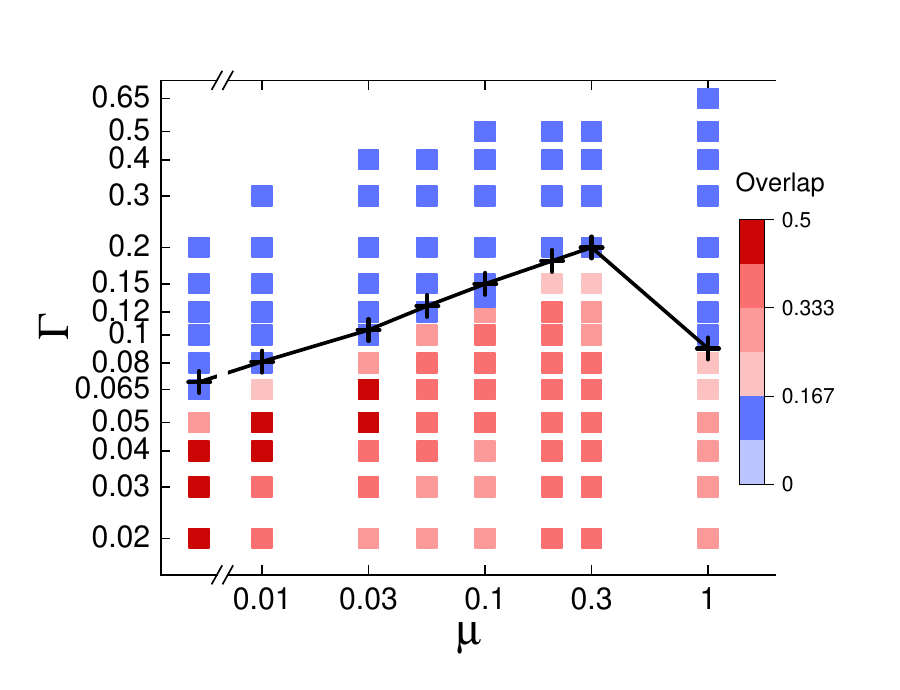}
\caption{Correlation between fast particles and large local packing fraction changes, quantified by the 
overlap between the top 10\% fastest particles and the top 10\% of particles with the largest local packing fraction changes
(see Methods for definitions). This figure provides a complementary view that support the dynamic regime boundaries shown in the state diagram (Fig.~3(a) in main text). The line with crosses marks the crossover strain $\Gamma_C(\mu)$.
}
\label{FigureS9}
\end{figure*}

\clearpage

\begin{figure*}
\centering
\includegraphics[width=16cm]{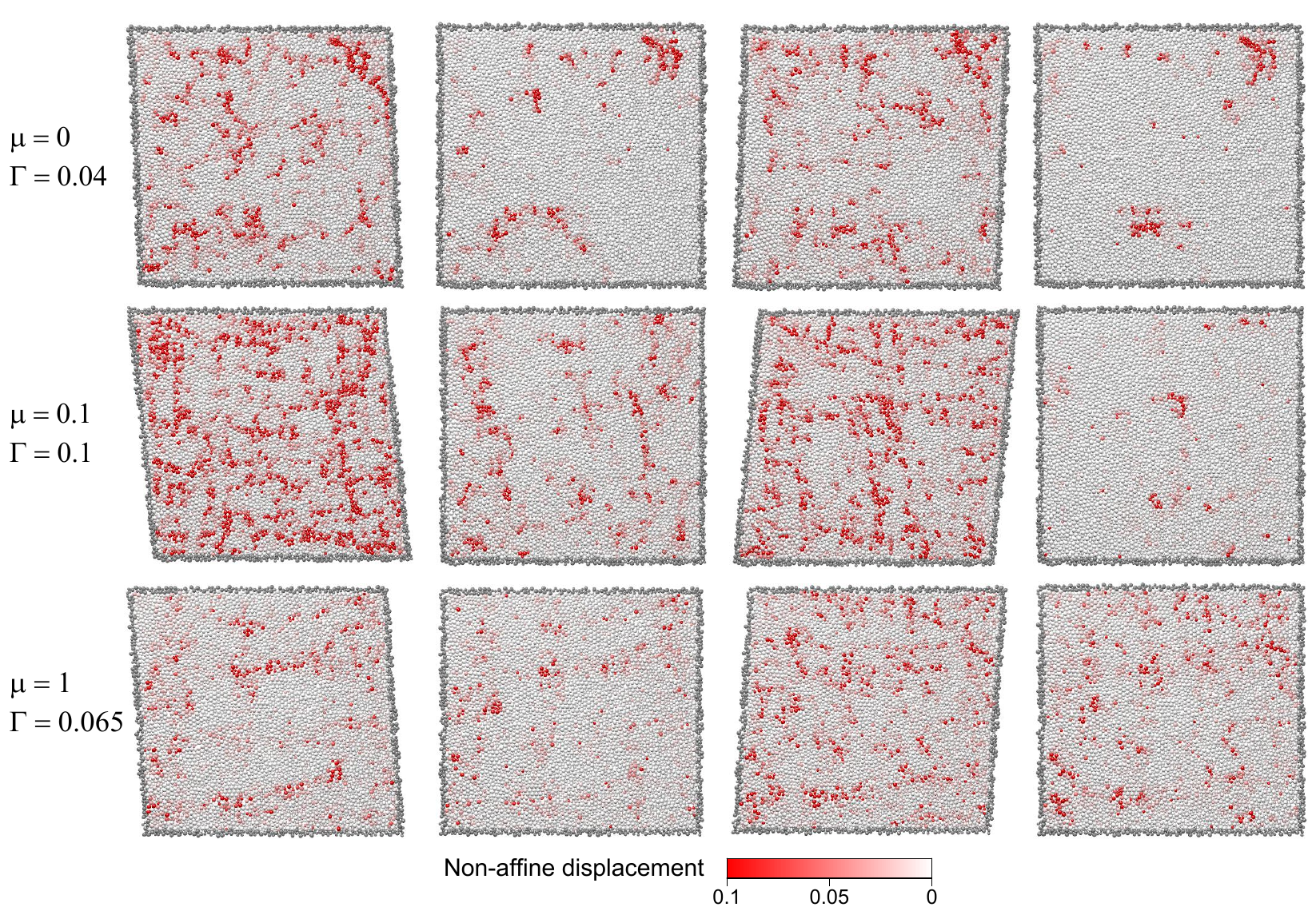}
\caption{Particle non-affine displacements $\vec{d}_i^2(\gamma)$ (see color bar) with respect to the origin $\gamma = 0$ within a full shear cycle $\gamma = 0 \rightarrow \Gamma \rightarrow 0 \rightarrow -\Gamma \rightarrow 0^*$ for three different systems (late-stage packings).
}
\label{FigureS10}
\end{figure*}

\end{document}